\documentclass[twocolumn]{aastex62}
\usepackage{graphicx}
\usepackage{lineno}
\usepackage{color}

\usepackage{epstopdf}
\usepackage{amsmath}

\def\ltsima{$\; \buildrel < \over \sim \;$}
\def\simlt{\lower.5ex\hbox{\ltsima}}
\def\gtsima{$\; \buildrel > \over \sim \;$}
\def\simgt{\lower.5ex\hbox{\gtsima}}

\newcommand\lsim{\mathrel{\rlap{\lower4pt\hbox{\hskip1pt$\sim$}}
\raise1pt\hbox{$<$}}}
\newcommand\gsim{\mathrel{\rlap{\lower4pt\hbox{\hskip1pt$\sim$}}
\raise1pt\hbox{$>$}}}

\allowdisplaybreaks
\usepackage{color}

 \def\fboundPrec{$13\%~$}
 
 \def\rcapCross{$0.002~$}
 
 \def\fGaia{$0.08$~}
\begin{document}

\title{Triples as Links between Binary Black Hole Mergers, their Electromagnetic Counterparts, and Galactic Black Holes } 

\author{Smadar Naoz}
\affiliation{Department of Physics and Astronomy, University of California, Los Angeles, CA 90095, USA}
\affiliation{Mani L. Bhaumik Institute for Theoretical Physics, Department of Physics and Astronomy, UCLA, Los Angeles, CA 90095, USA}
\author{Zolt\'an Haiman}
\affiliation{Institute of Science and Technology Austria (ISTA), Am Campus 1, Klosterneuburg, Austria}
\affiliation{Department of Astronomy, Mail Code 5246, Columbia University, New York, NY 10027, USA}
\affiliation{Department of Physics, Mail Code 5255, Columbia University, New York, NY 10027, USA}
\author{Eliot Quataert}
\affiliation{Department of Astrophysical Sciences, Princeton University, Princeton, NJ 08544, USA}
\author{Liz Holzknecht}
\affiliation{Department of Physics and Astronomy, University of California, Los Angeles, CA 90095, USA}
\affiliation{Mani L. Bhaumik Institute for Theoretical Physics, Department of Physics and Astronomy, UCLA, Los Angeles, CA 90095, USA}

\begin{abstract}

We propose a formation pathway linking black holes (BHs) observed in gravitational-wave (GW) mergers, wide BH–stellar systems uncovered by Gaia, and accreting low-mass X-ray binaries (LMXBs). In this scenario, a stellar-mass BH binary undergoes isolated binary evolution and merges while hosting a distant, dynamically unimportant tertiary stellar companion.  The tertiary becomes relevant only after the merger, when the remnant BH receives a GW recoil kick. Depending on the kick velocity and system configuration, the outcome can be: (i) a bright electromagnetic (EM) counterpart to the GW merger; (ii) an LMXB; (iii) a wide BH–stellar companion resembling the Gaia BH population; or (iv) an unbound, isolated BH. Modeling the three-body dynamics, we find that $\sim 0.02\%$ of LIGO–Virgo–KAGRA (LVK) mergers may be followed by an EM counterpart within $\sim$10 days, produced by tidal disruption of the star by the BH. The flare is likely brightest in the optical–UV and lasts days to weeks; in some cases, partial disruption causes recurring flares with a period of $\sim$2 months. We further estimate that this channel can produce $\sim 1–10\%$ of Gaia BH systems in the Milky Way. This scenario provides the first physically motivated link between GW sources, Gaia BHs, and some X-ray binaries, and predicts a rare but robust pathway for EM counterparts to binary BH mergers, potentially detectable in LVK's O5 run.
\end{abstract}

\section{Introduction}
Stellar-mass black holes (BHs) can be detected through multiple observational methods, each of which potentially probes different evolutionary pathways. The oldest method relies on electromagnetic (EM) signatures, most notably X-ray emission from accretion disks in X-ray binaries. These systems, discovered in the 1960s, provided the first indirect evidence for stellar-mass BHs \citep[e.g.,][]{Bolton72,Remillard+06,MacLeod+23}. 
More recently, the detection of gravitational waves (GWs) from merging binary BHs by the LIGO–Virgo–KAGRA (LVK) Collaboration has revolutionized our understanding of the BH mass distribution \citep[e.g.,][]{Abbott+16firstGW}, revealing a population of heavy BHs (up to $\sim 100$~M$_\odot$) without any known EM activity\footnote{A handful of LVK BHs have been suggested to coincide with EM flares in AGN~\citep{Graham+2023,Huang+25} but their brightness is difficult to explain~\citep{Tagawa+2024} and the significance of the spatial associations are controversial~\citep{Veronesi+25}.}.
The third avenue of detection, based on astrometric measurements of orbital motion, has been enabled by Gaia, which has opened a new observational window into detached BH binaries \citep{El-Badry+23GaiaH1}.   Stellar-mass BHs can also be detected by microlensing \citep[e.g.,][]{Agol+02,Lu+16,Lam+22,Sahu+22} or by combined radial velocity and photometric obeservations of non-accreting BHs in binaries \citep[e.g.,][]{Thompson+19,Liu+19BHRV,Clavel+21,Chawla+24}, but our focus in this paper is on the first three channels summarized above.

In addition to Gaia BH1 \citep{El-Badry+23GaiaH1}, two additional candidates have been identified in Gaia DR3 astrometric solutions \citep{GaiaCollab+21,GaiaCollab+23,El-Badry+23BH2,GaiaCollab+24BH3}, further demonstrating the potential of this technique to uncover detached BH binaries. However, the properties of Gaia BH1 in particular pose a significant evolutionary challenge: the orbital separation of the system lies in the range expected for the progenitor’s red super-giant phase, where the envelope would have engulfed the companion and likely led to a common-envelope phase \citep{El-Badry+23GaiaH1}. 

While each of these populations provides complementary information, the connections between them remain poorly understood. Do the merging BH binaries observed by LVK originate from the same progenitors as the X-ray binaries seen in the Galaxy? Are Gaia BHs an evolutionary link, or do they represent a distinct population? Reconciling the demographics and evolutionary pathways of these BH populations is a major open question, with implications for binary stellar evolution, supernova physics, and the formation of merging compact-object binaries. 

Recent efforts have begun to explore this question using both population synthesis and targeted observations, but a coherent picture has yet to emerge \citep[e.g.,][]{Zevin+21,Wiktorowicz+19,Fishbach+22,Fishbach+25}. 
For example, \citet{Zevin+21}  analyzed the second LIGO–Virgo catalog using population models for multiple BH binary formation pathways. They found that the diversity of observed mergers is best explained by a mixture of channels rather than a single formation mechanism producing more than $\sim 70\%$ of the detected population; they also highlighted how assumptions about natal spins and common-envelope evolution strongly affect inferred branching fractions. Similarly, \citet{Fishbach+22} compared BHs in X-ray binaries with those in GW-detected binary BHs. They showed that differences in masses between BHs in GW events and XRBs can largely be explained by observational selection effects and binary mass correlations, with natal BH kicks possibly also playing a role~\citep{Fishbach+25}. However, they also found a significant tension in the spin distributions, suggesting that while some binary BHs may have evolutionary histories similar to those of XRBs, others likely form through distinct channels.

\begin{figure}
  \begin{center} %\vspace{-1.2cm} %\hspace{0.2cm} 
\includegraphics[width=0.98\linewidth]{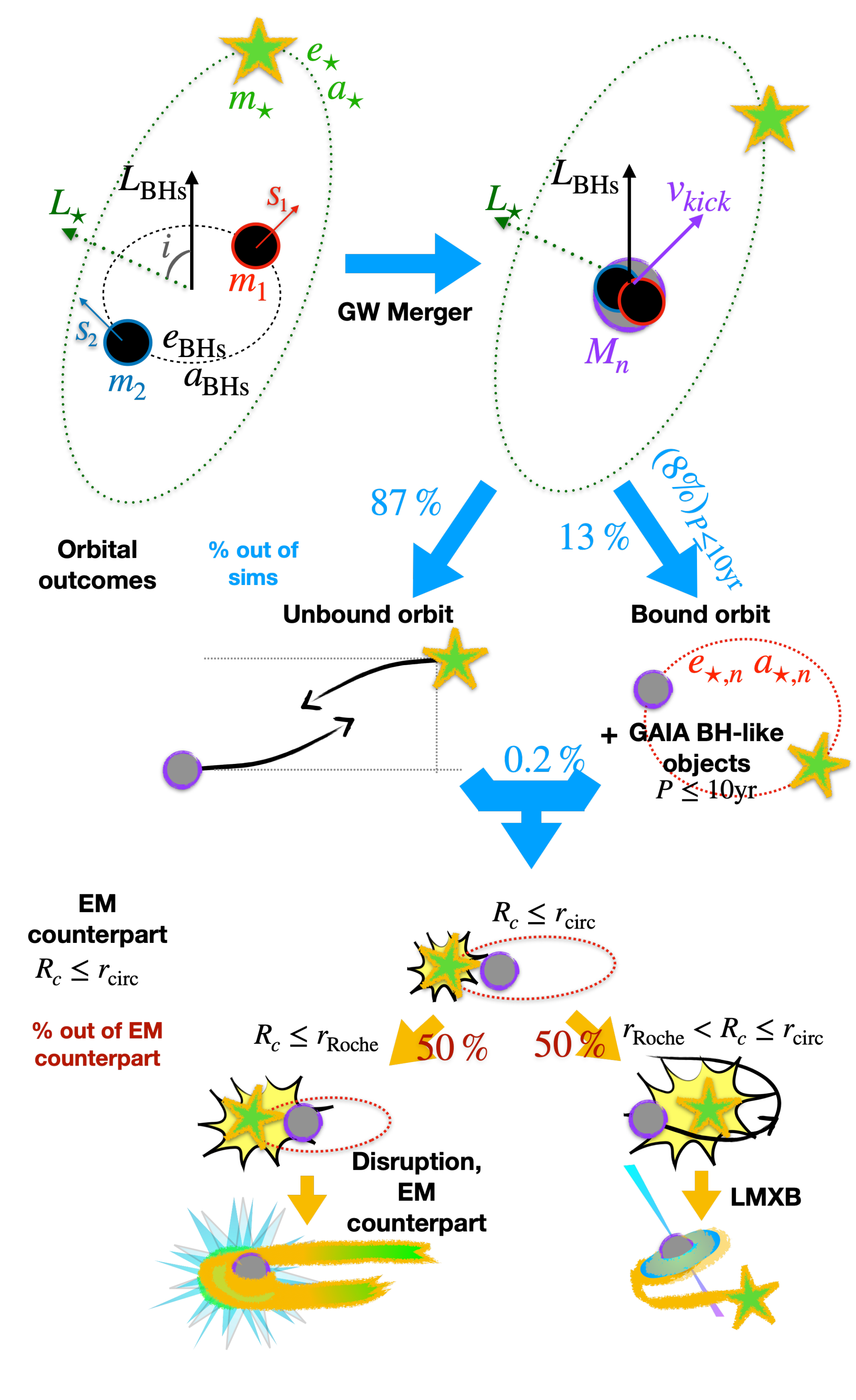}
  \end{center} %\vspace{-0.4cm} %\hspace{0.2cm}
  \vspace{-0.8cm}
  \caption{\upshape {\bf{An illustration of the system.}} The three-body system is composed of two stellar mass BHs and a distant main-sequence stellar tertiary. 
  The angle between the first (second) BH's spin, $S_1$ ($S_2$) and the inner binary angular momentum $L_{\rm BHs}$ is $i_{s1}$ ($i_{s2}$), and the angle in the plane of the orbit, 
  measured from the semimajor axis, is $\varphi_{1}$ ($\varphi_{2}$). Finally, the angle between the inner binary angular momentum and the recoil kick vector is $\alpha$. Post-recoil kick, there are two possible outcomes: unbound orbits (87\% of all systems) and bound orbits (13\% of all systems). We highlight that 8\% of all systems are bound with a period shorter than 10~years, which are Gaia-BH-like systems. %In some cases ($0.2\%$), 
  %the outcome of 
  In $0.2\%$ of all cases, either a bound or an unbound system results in an EM counterpart. These are designated by having the closest approach $R_c\leq r_{\rm circ}$, where $97.6\%$ (2.4\%) of all EM-bright sources are on a bound (unbound) orbit when they cross $r_{\rm circ}$. %About half of all the electromagnetic signature candidates systems cross the Roche radius, $r_{\rm Roche}$. Specifically, out of all those with radius smaller than the tidal radius, $13\%$ have $R_c\leq r_\star$, which may result in fast blue optical transients (FBOTs). 
 % About $13\%$ of those have their closest approach smaller than the star's radius, i.e., $R_c\leq r_\star$ resulting in a direct collision.
  %which may  result in fast blue optical transients (FBOTs).  
 % Additionally,  $37\%$ have $r_\star < R_c\leq r_{\rm Roche}$, which leads to the tidal disruption of the star, with an electromagnetic flare that is probably quite different from TDEs around a supermassive black hole. 
 About 50\% of these systems have closest approach smaller than $r_{\rm Roche}$, resulting in a prompt disruption event, and yielding an EM counterpart to the GW emission with a time delay of about $10$~days.  Finally, $50\%$ of the EM-signatures have $r_{\rm Roche} < R_c\leq r_{\rm circ}$, which may result in a low-mass X-ray binary (LMXB). Note that as stars evolve beyond the main sequence and become red giants, their radius expands, yielding a larger fraction of systems undergoing mass transfer events.  We discuss these probabilities in detail in \S~\ref{sec:EMrates}. 
  %Of which, 48.2\%  (0.8\%)  are on a bound (unbound) orbit when they cross $r_{\rm Roche}$, see Section \ref{sec:rates} for details. 
  %13% crossed rstar, 37% rstar > r > rRoche, 50% rroche > r > rcap
  }
  \label{fig:cartoon}
\vspace{0.4cm}
\end{figure}

In this {\it Letter}, we investigate whether the high prevalence of triples among massive stars  \citep[$\gtrsim 70\%$][]{Moe+17} can naturally connect the populations of BHs observed astrometrically, electromagnetically, and via GWs. In our scenario, when the inner BH binary in a hierarchical triple merges, the tertiary companion can remain bound, producing a wide BH binary detectable by Gaia, or in other cases, undergoes 
%unstable 
mass transfer, leading to an X-ray binary phase. In some cases, the tertiary companion is tidally disrupted, potentially producing luminous EM afterglows promptly following the merger.
This triple-mediated evolutionary pathway therefore provides a unified framework that can produce Gaia BH systems, X-ray binaries, and merging binary BHs, helping to explain the diversity and relative abundances of the observed BH populations.  It also provides a robust, albeit rare, channel for producing EM counterparts to binary BH mergers.  Fig.~\ref{fig:cartoon} illustrates the different channels we consider and their connections.

We stress that while hierarchical triples have been proposed as a formation channel for merging binary BHs in the field, their overall contribution to the LVK-detected merger rate is typically estimated to be small and suppressed by a factor of $\sim 5$–$30$ relative to the estimated detected rate \citep[e.g.,][]{Antonini+16,Antonini+17,Silsbee+17,Liu+19,Dorozsmai+24,Kummer+25}.  Although it is important to note that each of these studies has surveyed a limited part of the parameter space. Nonetheless, triple evolution may naturally account for certain systems that are otherwise difficult to explain, such as BH mergers in the mass gap, and other puzzles  \citep[e.g.,][]{Vigna-Gomez+21,Lu+21TripleGW,Martinez+22,Dorozsmai+25}. In this work, however, we focus on a complementary scenario: BH binaries that merge through isolated binary evolution, with a tertiary companion that does not play a significant dynamical role in driving the merger. In particular, the tertiary star's presence is relevant only after the inner binary merges, when the system receives a recoil kick. Our setup, therefore, explores the consequences of a passive tertiary in an otherwise binary-driven channel.   We assume throughout this paper that the distant tertiary companion remains bound during the formation of the two BHs in the inner binary, and focus on the dynamics that follow when the inner BH binary merges. The validity of this assumption is sensitive, of course, to the uncertain mass ejection and natal kick during BH formation.    Recent detections of X-ray binaries with distant companions imply that at least some distant companions remain bound during BH formation \citep[e.g.,][]{Burdge+24,Shariat+25xray}.

This {\it Letter} is organized as follows: We begin by describing the system setup in Section \ref{sec:setup}. % and how the kicks are implemented in Section \ref{sec:kicks}. 
In Section \ref{sec:outcomes}, we provide a clear explanation of the effects of a kick on stellar orbits. To test the scenario, we conduct a proof-of-concept population study in Section \ref{sec:pop}. We use this population to estimate the rate and detectability of EM signatures in Section \ref{sec:EMrates} and Gaia-BH detections in Section \ref{sec:Gaiarates}. Finally, we present our discussion and conclusions in Section \ref{sec:diss}.

\begin{figure*}
  \begin{center} %\vspace{-1.2cm} %\hspace{0.2cm} 
\includegraphics[width=\linewidth]{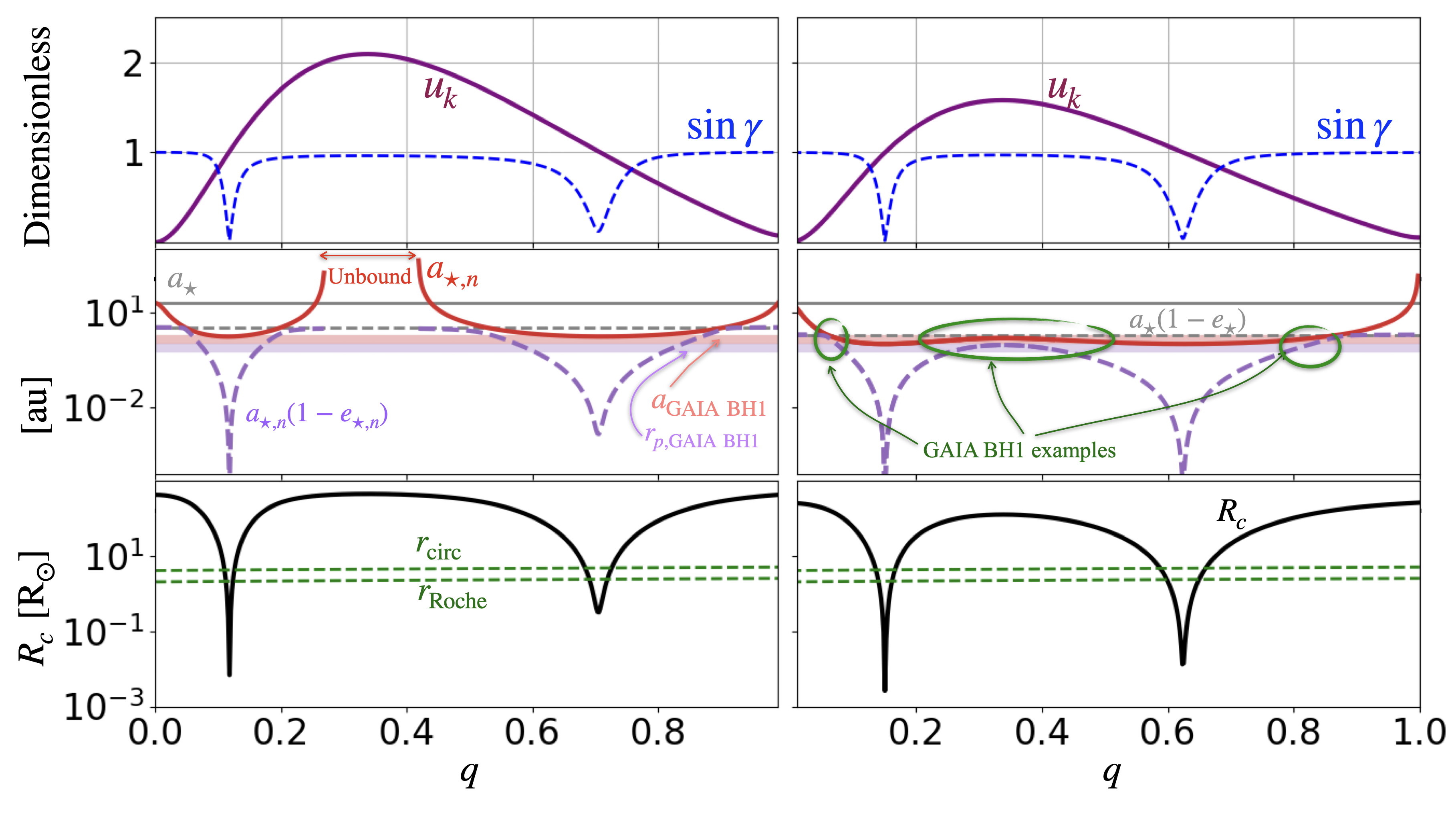}
  \end{center} %\vspace{-0.4cm} %\hspace{0.2cm}
  \caption{  \upshape {\bf{ Two examples of the orbital outcome of a recoil kick on an orbit}}. We consider a system of two low-spin BHs with $S_1=S_2=0.01$, and spin-orbit angles of $2^\circ$, $1^\circ$, $\Omega_{s1}=90^\circ$ and $\Omega_{s2}=270^\circ$. The mutual inclination is set to $0.1^\circ$. 
  Before the kick, the star has a semi-major axis of $20$~au and eccentricity $e_\star=0.9$ ({\it right column}), and $e_\star=0.83$ ({\it left column}). The argument of periapsis of the inner (outer) binary at the time of the merger is $300^\circ$ ($20^\circ$). These angles are relevant for the rotation of the various vectors to the invariant plane. See text for more details.   We assume that the kick took place when the star's phase was at $10^\circ$ ({\it right column}), and $5^\circ$ ({\it left column}). 
   Various post-merger quantities are shown as a function of the mass ratio of the two original BHs, $q$.
  {\bf The bottom row} shows the closest approach, Equation (\ref{eq:Rc}). Overplotted are the tidal 
  radius, $r_{\rm circ}$, and the Roche radius $r_{\rm Roche}$. We expect EM signatures for BHs that approach their stellar companion within these radii.  {\bf The middle row} depicts the post-kick semi major axis $a_{\star,n}$, solid red lines and pericenter $a_{\star,n}(1-e_{\star,n})$, dashed purple lines. Overplotted are the initial (pre-kick) semi-major axis and pericenter of the star, solid and dashed gray lines, respectively.  Lastly, Gaia-BH1's semi-major axis and pericenter ($a_{\rm Gaia~BH1}$, and $r_{p,\rm Gaia~BH1}=a_{\rm Gaia~BH1}(1-e_{\rm Gaia~BH1})$) are also overplotted, in pink and light purple lines \citep{El-Badry+23GaiaH1}.  {\bf The top row} shows the normalized velocity $u_k=v_{\rm kick}/v_r$, and $\sin\gamma$, where $\gamma$ is the angle between the radius vector and the post-kick velocity vector. The grid highlights that when $u_k=1$, $\gamma\to 0^\circ$, thus $R_c\to 0$.   Note that here a lower (higher) pre-kick eccentricity, while keeping all of the other parameters constant, yields a wider, more easily unbound (tighter) post-kick binary. The opposite trend is expected for a kick that takes place at apo-center, see Equations (\ref{eq:singamma2apo}) and (\ref{eq:ukperi}).    }
  \label{fig:OrbitEffectExample}
\vspace{1.4cm}
\end{figure*}

\section{System Set-Up and 
%Relevant 
Equations}\label{sec:setup}
\subsection{The Triple Configuration}
Throughout this work, we consider a stellar-mass BH binary, $m_1$ and $m_2$, where the mass ratio is $q\equiv m_1/m_2\leq 1$, where  ``1'' stands for the less massive component. 
%ZH: this may be nitpicking, but I think the standard is the opposite convention: "primary=1" and "secondary=2".
%SN: I have followed Lousto's formalizem - sorry.
The total mass of the binary is $M=m_1+m_2$, semi-major axis $a_{\rm BHs}$ and eccentricity $e_{\rm BH}$, associated with angular momentum $L_{\rm BHs}$. We define the spin-orbit angle of mass $m_j$ as $i_{sj}=\hat{h}\cdot\hat{S}_j$, where $\hat{h}$ is the unit vector along the BH's orbital angular momentum, and $j\in\{1,2\}$. The corresponding spherical coordinate angles are $\Omega_{sj}$ for each mass.   
Further, this binary is orbited by a star with a mass $m_\star=1$~M$_\odot$, and semi-major axis $a_\star$ and eccentricity $e_\star$. We note that while triples are observed to have a twin excess \citep{Shariat+251000}, the long lifetime of $1$~M$_\odot$ benefits the calculation.  In other words, the star is less likely to evolve before the BH binary merges. A full IMF calculation for the tertiary mass is beyond the scope of this paper. 
The frame of reference considered here is the inertial frame, for which the $z$-axis is parallel to the total angular momentum. After the BHs merge, they typically receive a recoil kick due to the anisotropic emission of GWs. 

\subsection{Impact on the Stellar Orbit}\label{sec:outcomes}

We adopt the \citet{Lousto+10,Lousto+12} fitting formulae, which provide the recoil kick velocity.
%within the plane of the binary BHs. 
%ZH: Wait... Lousto formulae work for arbitrary angles, and also give the out-of-plane kick velocity component. No?
%  Maybe delete "within the plane of the binary BHs" here.
%SN: yes, I used the general formulae
See Appendix \ref{sec:kicks} for the relevant set of equations. In this Letter, we focus on the merger channel driven by binary stellar evolution. This channel implies that the BH spins are aligned with its angular momentum. As a result, a well-known outcome emerges: the recoil kick is directed within the plane of the BH binary \citep[e.g.,][]{Lousto+12}. {Note that since natal kicks may induce misalignment, the recoil kick may be tens of degrees off-axis.  }
We discuss the effects of the spin-alignment approximation in Section \ref{sec:Gaiarates}. 

Assuming that the kick is instantaneous implies that the star and the new BH's separation does not change, i.e., $r=r_{\rm new}$ and that they now have a velocity vector: ${\bf v}_{\rm new} = {\bf v}_r-{\bf v}_{\rm kick}$, where ${\rm v}_r$ is the velocity vector of the outer binary just before the kick took place. The kick can either unbind the star-new BH system or alter its orbital configuration.

We can estimate the closest approach considering  gravitational focusing,
\begin{equation}\label{eq:GF}
    b^2 = R_c^2 + R_c\frac{2G(m_\star+M_{\rm new})}{v_{\rm new}^2} \ ,
\end{equation}
where $M_{\rm new}$ is the post-merger BH mass computed from \citet{Lousto+10}. 
This impact parameter is found using the post-kick velocity vector and separation:
\begin{equation}
    b=r\sin\gamma \quad {\rm where} \quad \sin\gamma = \frac{|{\bf v}_{\rm new}\times{\bf r}|}{|{\bf v}_{\rm new}||{\bf r}|} \ .
\end{equation}
Solving for $R_c$ we have:
\begin{equation}\label{eq:Rc}
    R_c =  \sqrt{b^2 + \frac{G^2(m_\star+M_{\rm new})^2}{v_{\rm new}^4}} -\frac{G(m_\star+M_{\rm new})}{v_{\rm new}^2} \ . 
\end{equation}
Clearly, when $b\ll G(m_\star+M_{\rm new})/v_{\rm new}^2$, the resulting $R_c$ will be small. This implies that for an arbitrary $r$, the post-kick binary velocity, ${\bf v}_{\rm new}$ should be close to parallel to the star's separation vector. Generally, the parameter $\sin\gamma$ can be written as: 
%Using cross product rules for ${\bf v}_{\rm new}\times {\bf r}$ we have:
\begin{equation}\label{eq:singamma2}
    \sin^2 \gamma = 1-\frac{({\bf v}_{\rm new}\cdot {\bf r})^2}{v_{\rm new}^2r^2} \ ,
\end{equation}
where ${\bf v}_{\rm new}\cdot {\bf r}=({\bf v}_{r}-{\bf v}_{\rm kick})\cdot {\bf r}$. 

To gain insight for the expected result, consider a situation where the recoil kick takes place when the star is at its apocenter (or pericenter), thus, ${\bf v}_{r}\cdot {\bf r}=0$. Therefore Equation (\ref{eq:singamma2}) takes the following form:
\begin{equation}\label{eq:singamma2apo}
    \sin^2 \gamma = 1-\frac{u_k^2\cos^2\alpha}{1+u_k^2-2u_k\cos\theta} \quad {\rm at~apo/peri-center} \ ,
\end{equation}
where we defined normalized velocity as  $u_k\equiv v_k/v_r$,  ${\bf v}_r\cdot {\bf v}_{\rm kick}=v_rv_{\rm kick}\cos\theta$, and ${\bf v}_{\rm kick}\cdot {\bf r}=rv_{\rm kick}\cos\alpha$.   Note that based on the geometry of the system, 
\begin{equation}
    \eta-\theta \leq \alpha \leq \theta+\eta \ ,
\end{equation}
where $\cos\eta = {\bf v}_r\cdot {\bf r}/(v_r r)$. At apo/peri-center $\eta=90^\circ$. Thus, in the apo/peri-center case, $\sin\alpha=\cos\theta$. 
Achieving small $R_c$ implies $\sin\gamma\to0$; thus, we solve Eq~(\ref{eq:singamma2apo}), in this case:
\begin{equation}\label{eq:ukperi}
    1+u_k^2\cos^2\theta -2u_k\cos\theta = 0 \quad {\rm at~apo/peri-center} \ .
\end{equation}
A straightforward solution occurs when \(\theta = 0^\circ\) and \(u_k = 1\). This means that the recoil velocity vector is directly parallel to the star's (pre-kick) velocity vector and has the same magnitude. In this scenario, we can expect \(\sin \gamma \to 0\), which implies a small value for \(R_c\). This situation also emphasizes the importance of the angle \(\theta\) in determining the orbital configuration after the kick. This behavior is depicted in Figure  \ref{fig:OrbitEffectExample}, top panel, where for most values of $q$, we have $\theta\approx0^\circ$. Considering two example systems, that demonstrate that when $u_k\to 1$, we have $\sin\gamma \to 0$, and as shown in the bottom panel, it results in $R_c\to 0$.

This example highlights that the mutual inclination between the two orbits is not a crucial factor after the kick. Rather, the orientation of the kick velocity in relation to the star's velocity is more important in determining the outcome of the orbit, particularly when the magnitudes of the two velocities are similar. Larger kick velocities, such as in the case when the spins are misaligned with respect to the BH binary's angular momentum, yield more unbound systems.

In the bottom row of Figure \ref{fig:OrbitEffectExample}, we compare $R_c$ to the tidal radius (top dashed line): 
\begin{equation}\label{eq:rcapt}
    r_{\rm circ} \sim 2 R_\star \left( \frac{M_{\rm new}}{m_\star}\right)^{1/3} \ .
\end{equation}
and to the Roche radius:
\begin{equation}\label{eq:rRoche}
    r_{\rm Roche} \sim  R_\star \left( \frac{M_{\rm new}}{m_\star}\right)^{1/3} \ ,
\end{equation}
(bottom dashed line). 
%Crossing 
A BH approaching the star inside either of these limits will likely result in an EM source.  Orbits with $R_c \lesssim r_{\rm Roche}$ will result in the tidal disruption of the star \citep[or, in some cases, a direct collision, e.g.,][]{Perets+16,Fragione+19,Rose+22,Yang+22,Ryu+22,Ryu+23,Ryu+24,Xin+24}.   Orbits with {\bf $r_{\rm Roche}\lsim R_c \lesssim r_{\rm circ}$} will lead to the orbit circularizing by tides, eventually producing a mass transferring system such as a low-mass X-ray binary.  We discuss these outcomes in more detail below. Note that $r_{\rm circ}$  is related to the tidal capture radius \citet{Press+77}, but in our case, most of the post-BH merger orbits are already bound, so there is no ``capture.''  The relevant radius is instead the pericenter distance inside of which tides lead to circularization on a timescale less than the evolutionary time of the star.  This is likely a factor of a few larger than the traditional tidal capture radius used in Equation \ref{eq:rcapt}~\citep[see, e.g.][]{Wu2018}, so our estimates of the fraction of the stars producing X-ray binaries are likely conservative.

\begin{figure*}
    \centering
    \includegraphics[width=\linewidth]{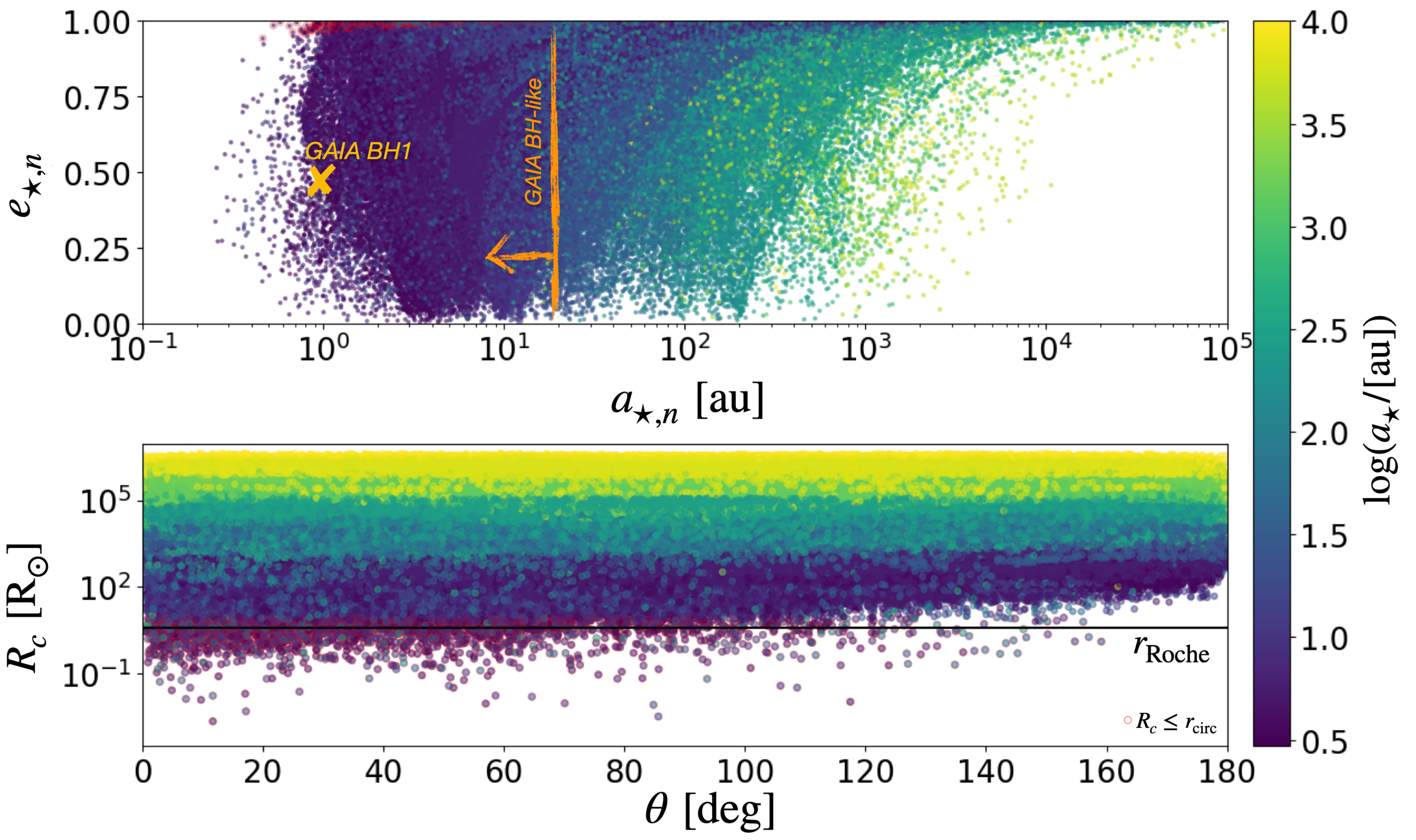}
    \caption{Post-kick orbital configurations of the proof-of-concept population. {\bf Top panel} presents the post-kick bound population ($13\%$ of all systems), and shows the eccentricity (y-axis) and semi-major axis (x-axis) of the star-BH orbit. Over-plotted are the orbital parameters of Gaia BH1 \citep[e.g.,][]{El-Badry+23GaiaH1} and the approximate part of the parameter space, potentially detectable by Gaia (defined by having a period up to 10 years and pericenter larger than $r_{\rm circ}$). {\bf Bottom panel} shows the closest approach (which is the pericenter for bound systems) of all of the systems in the simulation as a function of the angle between the recoil kick vector and the star's initial (pre-kick) 
    orbital velocity, i.e., $\theta = {\bf v}_r\cdot {\bf v}_{\rm kick}/(v_r v_{\rm kick})$. 
    Over-plotted is the Roche radius. The color code depicts the star's initial (pre-kick) semi-major axis.  The points with red edges are those that have $R_c\leq r_{\rm circ}$ and are therefore likely to produce EM emission.
        }
    \label{fig:population}
\end{figure*}

The orbital configuration of the bound binary can be estimated by assuming that the kick is instantaneous \citep[e.g.,][]{Kalogera00}
\begin{equation}
\label{eq:SMA1}
\frac{a_{\star,n}}{a_{\star,0}} = \frac{\beta (1-e_{1}\cos E_\star)}{2\beta - (1+e_{\star}\cos E_\star)(1+u_k^2-2u_k\cos\theta)} \ ,
\end{equation}
with 
\begin{equation}
\beta=\frac{m_1\star+M_{n}}{m_\star+M} \ ,
\end{equation}
where the normalized velocity and the angle $\theta$ are defined above.  The post-kick semi-major axis can shrink (expand) if $\beta$ is larger (smaller) than $\beta>1+u_k^2-2u_k\cos\theta$ \citep[][]{Lu+19}.
 The post-kick BH-stellar orbital eccentricity is given by \citep{Lu+19}
\begin{equation}\label{eq:e1n}
e_{\star,n}^2=1-\frac{|{\bf r}\times ( {\bf v}_{r}+{\bf v}_{\rm kick}) |^2 }{ a_{\star,n} G(m_\star+M_{\rm new} )} \ .
\end{equation}
Note that ${\bf v}_{\rm kick}$ is calculated in the plane of the inner binary, while ${\bf v}_{r}$ is defined in the plane of the outer orbit. We thus rotate all vectors to the invariable plane, defined such that the $z$-axis is parallel to the total angular momentum.  We use the pre-kick relevant angles. Specifically, the relevant angles are the angles between each orbit's angular momenta and the total angular momentum, $i_1$ and $i_2$ for the inner and outer orbits, respectively, and the arguments of periapsis of the inner and outer orbits. In such a frame of reference, the difference between the inner and outer longitude of ascending nodes is $\pi$ \citep[e.g.,][]{Naoz+13sec}. 

Figure \ref{fig:OrbitEffectExample} illustrates the post-kick semi-major axis and eccentricity of the previously discussed example systems, comparing them to their pre-kick values (see labels). The left side shows an example where part of the parameter space resulted in the system becoming unbound. In contrast, the right column presents a scenario in which a newly formed BH and the star remained bound, leading to a contraction of the post-kick semi-major axis across a broad range of the parameter space. Overlaid on the figure are the Gaia-BH1 orbital parameters from \citet{El-Badry+23GaiaH1}. This example demonstrates that such processes can naturally lead to the formation of a Gaia-BH1 configuration, as well as other configurations of bound systems.

\section{A population study}\label{sec:pop}
As a proof of concept, we focus on a population of isolated BH binaries that will ultimately become LVK sources, and add a stellar tertiary to them.  The isolated binary channel  involves a common-envelope, stable mass transfer phase, or chemically homogeneous evolution \citep[e.g.,][]{Belczynski+02,Belczynski+07,Dominik+12,Dominik+15,Stevenson+17,deMink+16,Mandel+16,Gallegos-Garcia+21}. We thus adopt binary BH orbital configurations that isolated binary population synthesis work suggests will merge. Specifically, we are motivated by the angular momentum constraints from Figure 13 in \citet{Kruckow+18}, which represent the orbital configurations of two BHs, right after the formation of the second BH. Specifically, we choose the initial BH binary eccentricity from a uniform distribution {between 0 -1,} and apply the angular momentum constraints on the binary's semi-major axis. Additionally, the population is limited to a merger time of  $\leq 10$~Gyr  (which reduces the population by a factor of 2). {Note that an eccentric BH binary torques the star's inclination via the iEKL; at the time of merger, the BH binary dissipates its eccentricity via GW emission.   }  

For simplicity, the BH masses are adopted from a uniform distribution, each between $9-100$~M$_\odot$. {We note that the choice of mass distributions here may affect the resulting rates at the order of $\sim \pm 10-20\%$, as tested via toy models adopting a double power law, with Gaussian peaks for the LVK mass distribution, following \citet[][]{Callister+24} and \citet{LVKO4results}.  These tests are omitted here to avoid clutter. Given these and other uncertainties, we reserve a more detailed analysis for future endeavors. }

For the outer orbit, we are motivated by triple conditions that include post-main-sequence stellar evolution for massive stars \citep[e.g.,][]{Naoz+16,Stephan+19,Vigna-Gomez+21,Kummer+25,Shariat+25xray}. We thus choose the star's semimajor axis\footnote{{Note that the aforementioned studies yield wide tertiary orbital separations. Tighter configurations may increase the rates of the various outcomes.}} from a log-normal distribution between [3-100]~au, and a thermal distribution for the eccentricity \citep[e.g.,][]{Shariat+251000}. Note that this approach assumes zero (or very small) BH natal velocity~\citep[e.g.,][]{Naoz+16,Shariat+25xray}. Lastly, the mutual inclination between the inner and outer binary is chosen from an isotropic distribution (uniform in $\cos i$); the arguments of the inner and outer perihelions are chosen from a uniform distribution from $0-2\pi$. 

{The setup of the system is such that in most parts of the parameter space, the BH binary torques the star rather than the star influencing the dynamics of the BH binary. This can yield a preferred orientation for the star’s orbital plane \citep[e.g.,][]{Naoz+17,Hansen+20}. To find the possible effect of the star's orientation, we} solve the hierarchical triple-body equation of motion up to the hexadecapole level of approximation. There are two main reasons for the usefulness of this approximation beyond the octupole level. The first is that it allows us to integrate inner binaries with comparable masses. The second reason relates to the fact that in some of the systems, the ratio between the period of the outer orbit and the timescale of the quadrupole-level eccentric Kozai-Lidov (EKL) cycles is comparable to or larger than the strength of the octupole term. It was demonstrated that in such systems, the octupole level is insufficient, and the next-level approximation, i.e., hexadecapole, allows for a more accurate description of the dynamics  \citep[e.g.,][]{Soderhjelm75,Cuk+04,Luo+16,Will17,Will21,Tremaine23,Klein+24}. Additionally, we include the 1st post-Newtonian (1PN) precession of the inner and outer orbit, which can either suppress or excite eccentricity oscillations for either orbits \citep[e.g.,][]{Naoz+13GR,Naoz+17,Will17,Lim+20}. 
We neglect the 1.5PN and 2PN terms, which would slow down the calculation and only impact spin evolution and not significantly change the orbital precession.
On the other hand, we add the 2.5PN terms, which induce orbital shrinking and circularization via GW emission according to \citep[e.g.,][]{Peters+63,Peters64}.

The initial conditions, as well as all systems throughout their evolution, satisfy the dynamical stability requirement:
\begin{equation}
\epsilon = \frac{a_{\rm BH}}{a_\star} \frac{e_\star}{1 - e_\star^2} \leq 0.1 \ ,
\end{equation}
where $\epsilon$ is the dimensionless parameter that appears as the prefactor of the octupole-level term in the hierarchical three-body Hamiltonian \citep[e.g.,][]{LN11,Naoz+13sec}. Although various alternative stability criteria exist in the literature \citep[e.g.,][]{Mar10, Mushkin+20,Vynatheya+22,Tory+22,Zhang+23}, the $\epsilon$ criterion has been shown to be broadly consistent with many of these \citep[e.g.,][]{Naoz+14}. Furthermore, while the inclusion of higher-order terms, such as the hexadecapole-level approximation, can extend the validity of secular dynamics to more compact systems \citep[e.g.,][]{Will17}, the $\epsilon \leq 0.1$ threshold remains a conservative and robust choice.

Since each binary BH is chosen such that it will merge within 10~Gyrs, we explore the orbital configuration of the stellar tertiary in random intervals. Specifically, we sample the system 1000 times during the integration lifetime. The latter is set to explore the full dynamical extent of the system. Because the mass of the star is small compared to that of the inner binary, the inner binary can also torque the star, known as the inverse Eccentric Kozai Lidov (iEKL) mechanism \citep[e.g.,][]{Naoz+17,Zanardi+17,Vinson+18,Naoz+20}. For the entire population, we find that the iEKL quadrapole-level timescale for the population is much longer than the quadrapole-level timescale to torque the inner binary, i.e., 
\begin{equation}\label{eq:tquad}
t_{\rm quad}\sim \frac{16}{15}\frac{ a_{\star}^3 (1-e_{\star}^2)^{3/2}\sqrt{m_1+m_2}}{ a_{\rm BH}^{3/2} m_\star \sqrt{G}} \ .
\end{equation}
Thus, to allow sufficient simulation time for either the BH binary or the stellar orbital parameters to evolve dynamically, we adopt the descent timescale, $t_{\rm descent}$. This quantity describes the time required to reach extreme eccentricity via higher-order approximations of the Hamiltonian. Recently, \citet{Weldon+24} found an analytical expression for this timescale, i.e.,:
\begin{equation}\label{eq:tdec}
     t_{\rm descent} = t_{\text{quad}}  + \Upsilon t_{\rm oct} \eta(r_{p,\rm min}) \ , 
\end{equation}
where $t_{\rm quad}$ is defined in Equation (\ref{eq:tquad}), 
%\begin{equation}\label{eq:tquad}
%t_{\rm quad}\sim \frac{16}{15}\frac{ a_{\rm out}^3 (1-e_{\rm out}^2)^{3/2}\sqrt{m_1+m_2}}{ a_{\rm in}^{3/2} m_3 \sqrt{G}} \ ,
%\end{equation}
 and
\begin{equation}
    t_{\rm oct}=\frac{64}{15} G^{-1/2}\frac{a_{\star}^4}{a_{\rm BH}^{5/2}} \frac{(m_1+m_2)^{3/2}}{(m_1-m_2)m_\star}  \frac{(1-e_{\star}^2)^{5/2}}{e_{\star}} \ .
\end{equation}
We thus integrate each system to $t_{\rm age} =$min$(t_{\rm descent},1~{\rm Gyr})$ or until the inner binary merges. We discarded systems that became unstable during the evolution ($2\%$ of all systems), because the secular equations do not describe the full dynamics for these systems \citep[e.g.,][]{Naoz+17}.   

Lastly, at each output timestep, we adopt an eccentric anomaly for the star chosen from a uniform distribution. The BHs' spins remain aligned with $i_{s1}=2^\circ$ and $i_{s2}=2^\circ$, and $\Omega_{s1}$ and $\Omega_{s1}$ are chosen from a uniform distribution. 

Figure \ref{fig:population} depicts the results of this proof-of-concept population study. The top panel shows the post-kick bound systems' semi-major axis and eccentricity (representing \fboundPrec of the systems). 
The bottom panel shows the closest approach of all systems as a function of the angle $\theta$, which is the angle between the recoil kick velocity and the star's velocity vector just before the kick took place. Red circles indicate systems that have a closest approach (peri-center in the case of bound systems) smaller than $r_{\rm circ}$, from Equation \ref{eq:rcapt}.  
The color code represents the 
star's initial
%initial star's 
semi-major axis. 

As demonstrated in this example, Gaia-BH-like systems naturally form post-kick, and they happen for systems with typical initial semi-major axes of $\sim 12$~au. However, there is a long tail of initial separations that can produce Gaia-BH-like systems, with the widest one being $\sim 6\times 10^3$~au. Moreover, this channel produces a wide range of detached BH-star systems, reaching up to $10^4$~au in separation. 

The initial separation of the star is a key factor in determining the post-kick configuration, both for Gaia-BH-like systems and for those with $R_c \lesssim  r_{\rm circ}$. As highlighted in the bottom panel of Figure \ref{fig:population}, very wide ($\sim 10^4$~au) pre-kick stars are less likely to end up with a small $R_c$. Specifically, those that cross $r_{\rm circ}$ have, on average, a pre-kick semi-major axis of $13$~au,
but again with a long tail extending up to $a_{\star}\sim 10^3$~au. 

%{\sn{Note: I tested the distribution here, it's not very illuminating...it almost fits with Generalized Extreme Value (GEV) or Generalized Pareto. Which has a sharp peak and long tail in $\log a_\star$, but I'm not sure it tells us anything apart from the fact that sometimes we can be lucky.  }}

\begin{figure}
    \centering
    \includegraphics[width=
    \linewidth]{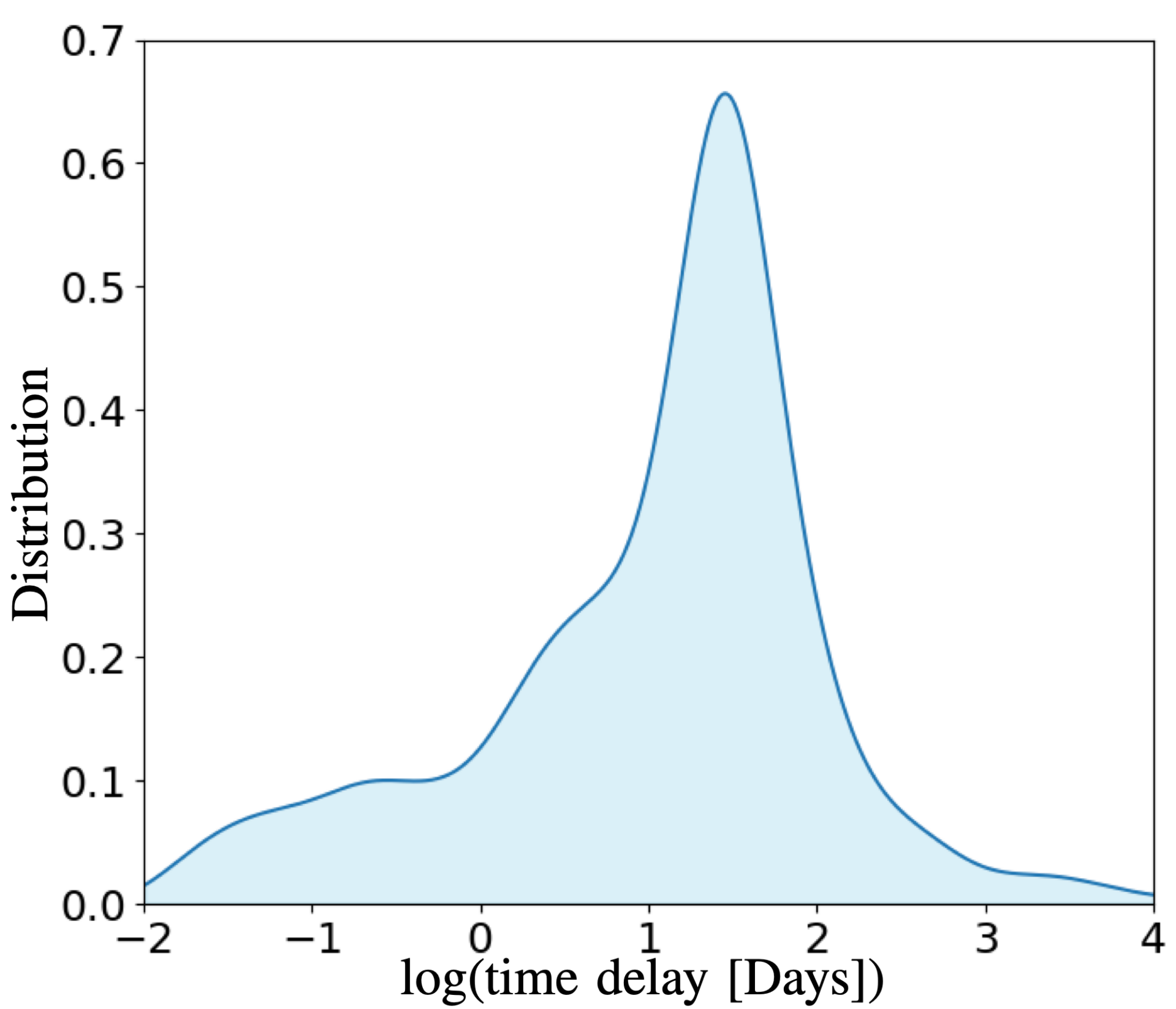}
    \caption{{\bf Time delay distribution for the EM counterpart}, assumed to be the time between the merger of the inner BH binary and when the BH remnant -- star separation shrinks below
    $r_{\rm circ}$.  
    Both bound ($87\%$ of all systems with $R_c\leq r_{\rm circ}$) and unbound systems, as well as those that crossed the star's Roche limit ($\sim 50\%$ out of all systems with $R_c\leq r_{\rm circ}$), exhibit the same distribution. The average of this distribution is $\sim 10$~days. We note that the fallback time of the bound debris onto the BH is on the order of a day, or less, see Equation (\ref{eq:TDEFB}). }
    \label{fig:TimeDelay}\vspace{0.5cm}
\end{figure}

\section{Rates and Detectability}\label{sec:rates}
\subsection{General considerations}

Rate estimates involve considerable uncertainty and should therefore be taken with a grain of salt. Here, we aim to estimate the order of magnitude contribution of this channel to EM signatures and Gaia-BH-like systems. We begin by assuming that the BH binaries produced by triples as modeled here are representative of LVK detections, with a rate estimated as $\Gamma_{\rm LVK}\sim 30$~Gpc$^{-3}$~yr$^{-1}$ \citep[e.g.,][]{LVK+23BBHRate}. We note that there are many different proposed channels for LVK mergers and that this assumption by itself is very strong. Other ideas in the literature suggest that dynamical scattering in globular clusters can produce a significant fraction of LVK sources \citep[e.g.,][]{Rodriguez+16,Rodriguez+18,Samsing18}, as well as other dynamical channels, such as in galactic nuclei \citep[e.g.,][]{OLeary+09,Hoang+18,Petrovich+17}, fly-by interactions in the field \citep[e.g.,][]{Michaely+20}, hierarchical triples \citep[e.g.,][]{Silsbee+17,Kummer+25}, or in AGN disks \citep[e.g.,][]{Tagawa+20,Samsing+22}. Regardless, the isolated binary channel is likely to also be significant \citep[e.g.,][]{Nutzman+04,deMink+16,Mandel+16,Belczynski+16,Marchant+16,Breivik+16}.  Therefore, the rate below can be normalized by the contribution of field binaries to LVK sources.   

Taking the field binaries as the main contributions, we remind the reader that 
our initial condition setup of the binary, as described in Section \ref{sec:pop}, represents about $50\%$ of the BH orbital configuration that will lead to mergers\footnote{As mentioned in Section \ref{sec:pop}, we select a BH binary population with a merger timescale shorter than 10 Gyr, which is effectively half of the population synthesis considered by  \citet{Kruckow+18}. }. In other words, $f_{\rm BH}\sim 0.5$. Note that we start with the LVK merger sample, but half of the mergers took too long for the tertiary $1$~M$_\odot$ star to still be around. Notably, about $70\%$ of massive binaries are in a triple configuration \citep[e.g.,][]{Moe+17,Offner+23}. However, since $\sim$solar mass companions can be missed, this is {may be} a lower limit, and we assume for simplicity that all massive binaries have a tertiary\footnote{{ Note that the fraction of surviving BH binaries with tertiaries remains uncertain. Specifically, while the observational constraints on unbound tertiaries to supernova counterparts are low \citep[e.g.,][]{Barboza+24}, the connection between these systems and LVK sources is yet to be established.}}. 
Furthermore, the fraction of these triples that host a $\sim 1$~M$_\odot$ star is $f_{1}\sim 0.2$, estimated from the tertiary mass-ratio distributions in triples in \citet{Shariat+251000}. While a more massive tertiary would increase the predicted fractions, its shorter main-sequence lifetime could limit the time available for this process to occur. A detailed study of the tertiary initial mass function is left for future work.

With these two fractions, we can proceed to estimate the probabilities of the different outcomes, as well as their EM detectability.

\subsection{EM signatures}\label{sec:EMrates}
Here we divide the discussion into systems that crossed the tidal radius $r_{\rm circ}$ ($f_{\rm capt}\sim$ \rcapCross out of all systems), and the $\sim 50\%$ of that subset that also cross the Roche radius and so the star is disrupted. Therefore, we find:
\begin{eqnarray}
    \Gamma_{\rm capt} &\sim & \Gamma_{\rm LVK} \times f_{\rm BH} \times f_1 \times  f_{\rm capt} \sim 0.006~{\rm Gpc}^{-3}~{\rm yr}^{-1} \\
     \Gamma_{\rm Roche} &\sim & \Gamma_{\rm LVK} \times f_{\rm BH} \times f_1 \times \frac{f_{\rm capt}}{2} \sim 0.003~{\rm Gpc}^{-3}~{\rm yr}^{-1} \ .
\end{eqnarray}
Since no reliable EM signature has been observed so far in LVK \citep[e.g.][]{Veronesi+25}, naturally, we expect a less than $1\%$ event rate, which agrees with the above estimates. Specifically, $0.02\%$ ($0.01\%$) of LVK BH binary mergers may be accompanied by EM counterparts, crossing $r_{\rm circ}$ ($R_{\rm Roche}$). If the future LVK campaigns O5 or O6 yield more than a total of 5000 events, we predict the possible detection of an EM counterpart associated with a BH merger. 

An important question to consider is: what is the time delay between a GW event and the crossing of $r_{\rm circ}$ and $R_{\rm Roche}$? In the triple scenario considered here, this is a lower limit on the time delay between an EM counterpart and the binary BH merger that preceded it.   Figure \ref{fig:TimeDelay} illustrates the distribution of time delays, indicating an average delay of about $10$~days between the merger and its EM counterpart.  All possible combinations of outcomes, whether the tertiary star is bound or unbound, and whether it crosses $r_{\rm circ}$ or $R_{\rm Roche}$, exhibit the same distribution. 

There are two distinct classes of EM sources that our scenario can produce.   Stars with $R_c \lesssim r_{\rm Roche}$ will undergo tidal disruption with a bright, relatively prompt EM flare, delayed relative to the GW event by the amount shown in Figure~\ref{fig:TimeDelay}.   By contrast, stars with $r_{\rm Roche} \lesssim R_c \lesssim r_{\rm circ}$ will slowly circularize by tides and eventually produce a mass-transferring binary, resembling a long-lived Galactic X-ray binary more than a prompt EM flare.  The time delay between the GW event and the onset of EM emission from mass transfer will typically be set by the stellar evolution timescale of the tertiary and so will be of order Gyr.   In this case, it will of course not be possible to temporally correlate the GW and EM signals, and the EM signal is likely to be far too faint to detect beyond nearby galaxies (i.e., individual XRBs cannot be seen at cosmological distances). 
%ZH: I am not sure if the above is true. There is definitely a catalog of extragalactic XRBs by Chandra, although they are probably low-z nearby galaxies - and maybe all high-mass XRBs ?
%SN: I thought that extragalactic XRB are all low z. We estimate the for high z we often think that the XRBs contibute to a background. See: https://ui.adsabs.harvard.edu/abs/2013ApJ...776L..31F/abstract
%EQ:   yes, there are extragalactic XRBs but at even moderate z it's the total luminosity of an XRB population that is detected not individual objects; the latter is only possible relatively locally.  that's sort of the point i was getting at ...
%ZH2: OK, good, yes, thanks.  I nitpicked and changed the wording to avoid "extragalactic" and instead say "too faint beyond nearby galaxies".    The Chandra catalog does contain many extragalactic XRBs but resolved only out to ~130 Mpc (https://ui.adsabs.harvard.edu/abs/2021PASJ...73.1315I/abstract)
%SN2: OK
We thus focus our discussion here on the regime of $R_c \lesssim r_{\rm Roche}$ which has the potential to produce a prompt detectable EM counterpart to cosmological binary BH mergers detected in GWs \citep[e.g.,][]{Perets+16,Fragione+19,Yang+22,Xin+24}.

In the regime $R_c \lesssim r_{\rm Roche}$, the primary outcome is an EM flare associated with the tidal disruption of the star by the newly formed BH.  The fallback timescale,  the time it takes the bound debris of the disrupted star to fall back onto the BH, is given roughly by
\begin{eqnarray}\label{eq:TDEFB}
   t_{\rm FB} &\approx&  2 \pi \frac{r_{\rm Roche}^3}{r_\star^{3/2}}\frac{1}{\sqrt{GM_{\rm new}}} = 4 \pi \frac{r_\star^{3/2}}{m_\star}\sqrt{\frac{M_{\rm new}}{G}} \ , \\
   &\approx & 1.6~{\rm days} \left(\frac{r_{\star}}{{\rm R}_\odot}\right)^{3/2} \left(\frac{m_\star}{{\rm M}_\odot}\right)^{-1}\left(\frac{M_{\rm new}}{50~{\rm M}_\odot}\right)^{1/2} \nonumber \ .
\end{eqnarray}
\citet{Coughlin+22} and \citet{Bandopadhyay+24} found that, for disruptions around a supermassive black hole, the fallback time may be shorter by an order of magnitude than this estimate, with little dependence on stellar mass for a main-sequence star. Therefore, equation \ref{eq:TDEFB} is likely an upper limit on the fallback time.   Equation \ref{eq:TDEFB} thus shows that the bound debris of the disrupted star will fall back to the BH within a day or so, much shorter than the time delay in Figure \ref{fig:TimeDelay} set by the time it takes the star and BH to interact after the GW merger.
%(numerical simulations suggest that the fallback time is perhaps $\sim 10 \times$ shorter than the estimate in equation \ref{eq:TDEFB}, reinforcing this conclusion; \citealt{Bandopadhyay2024}).  
The delay between the GW source and its EM counterpart will thus largely be set by the timescale given in Figure \ref{fig:TimeDelay}.

%About half of the EM-signature candidates undergo disruption with $R_c\leq r_{\rm Roche}$. 
%13% smaller than r_star and 37% between r_star and r_roche

Observations of TDEs by supermassive black holes show a rich phenomenology across the EM spectrum.  A small fraction of sources show bright non-thermal X-ray and gamma-ray emission associated with a relativistic jet, but such prominent jet emission is much rarer in TDEs than in active galactic nuclei or X-ray binaries \citep{Komossa+03}.   Instead the dominant signature of TDEs is thermal optical-UV-X-ray emission and radio emission produced by outflows interacting with the ISM \citep{Gezari2021,Cendes2024}.  

The key difference between the TDEs by solar mass BHs considered here and those associated with supermassive BHs is that the shorter fallback time and lower BH mass imply that the fallback rate is highly super-Eddington.   This is likely to suppress bright high-energy emission from the vicinity of the BH because it is enshrouded in the optically thick super-Eddington envelope (except perhaps for very favorable viewing angles down the spin axis of the system or if a jet escapes the optically thick envelope).   In this case, the most likely robust EM counterpart is a thermal optical-UV flare powered by the super-Eddington outflow, with luminosities of up to $\sim 10^{44} $ erg s$^{-1}$ and durations of days-weeks (e.g., \citealt{Kremer+19,Kremer+21}).  The sources will likely resemble the optical-UV emission in fast blue optical transients such as 2018cow \citep{Margutti2019, Perley2019}, which are also interpreted as TDEs or mergers between stars and stellar-mass compact objects or intermediate-mass BHs \citep{Metzger2022,Linial2024,Tsuna+25}.    

In a fraction of the TDE cases considered here with $r_{\rm Roche} \lesssim R_c \lesssim 2 r_{\rm Roche}$, the star will only be partially disrupted at pericenter likely leading to a periodic optical-UV transient with a period of about 2 months (set by Figure \ref{fig:TimeDelay}); this is analogous to partial tidal disruption candidates by supermassive black holes observed in galactic nuclei (e.g., \citealt{Payne2021}).   

Finally, we note that during the red giant phase, the probability of tidal disruption is significantly higher because of the larger stellar radius.  However, a solar mass star only spends $\sim 1 \%$ of its lifetime with a radius $\gtrsim 10 R_\odot$, so the red giant phase likely does not enhance the overall rate of EM counterparts significantly.   The fallback time for red giants will also be longer (eq. \ref{eq:TDEFB}), leading to a longer delay between the EM and GW sources and likely a fainter EM counterpart because of the lower fallback rates.

\subsection{Gaia-BH-like systems}\label{sec:Gaiarates}

To estimate the efficiency of this channel in producing Gaia-BH-like systems, we begin by estimating the total number of LVK sources in a galaxy, assuming that galaxies with masses comparable to the Milky-Way (MW) produce a significant fraction of the LVK rate (if they do not, the estimates below of Gaia-BH-like systems in the MW are upper limits). {We note that the LVK binaries are often thought to have originated from a low metallicity environment \citep[e.g.,][]{Dominik+12,Rodriguez+15,Fishbach+21}. However, the strong metallicity dependence of BH binary formation is still uncertain \citep[e.g.,][]{vanSon+25}. Thus, estimating the galactic BHs from LVK mergers is done here just to provide an order of magnitude sense of the number of galactic BHs where post merger recoil could be important for setting the systems properties. }

Given the number densities of massive galaxies $\sim 0.01-0.001$~Mpc$^{-3}$ \citep[e.g.,][]{Conselice+16}, over a Hubble time, the expected number of BH mergers per galaxy is $N_{\rm BBH}\sim 3\times 10^{4-5}$. Thus, to estimate the number of Gaia-like systems, we multiply this number by $f_{\rm BH} \times f_1$, i.e., the fraction that our initial conditions represent from the total simulated binary BH population. 
Then, we multiply this by $f_{P_{\leq10}}\sim$\fGaia, which is the fraction of systems out of all runs that remain bound after the kick, have a period smaller than $10$~years, and have a pericenter larger than $r_{\rm circ}$. Thus, the expected Gaia-BH-candidates from this channel are
\begin{equation}
    N_{\rm Gaia-BH}\sim N_{\rm BBH}\times f_{\rm BH} \times f_1 \times f_{P_{\leq10}} \sim  240-2400 \ .
\end{equation}
The total number of Gaia-BH-like systems in our Milky Way is estimated to be about 20,000 \citep{Nagarajan+25}, thus, this channel may contribute to 1-10\% of Gaia-BH-like systems. In our models, these Gaia-BH-like populations have an average separation of $\sim 7$~au and eccentricity of $0.6$. This estimate does not depend on the spin of the BHs, but it is sensitive to the alignment of the BHs, where misaligned BHs will result in even fewer post-kick bound systems. {A small misalignment for the less (more) massive BH $\lesssim $ 10 (5)~deg, has a negligible effect on the results.} While the binary merger channel naturally results in spin alignment, in some cases, where the star is on a tight orbit, it torques the inner, BH binary orbit, which may result in misalignment \citep[e.g.][]{Liu+18}.  We reserve this part of the investigation for future work.  

%We estimate that about $13\%$ of all systems remain bound, with an average separation of $\sim 594$~au and eccentricity of $0.6$.  Thus, we predict a large additional population of Gaia-BH systems with a wide range of separations.   
Overall, we estimate that about $13\%$ of all systems remain bound, with an average separation of $\sim 594$~au and eccentricity of $0.6$. Of these, about $60\%$ have a period of less than ten years, yielding a Gaia-BH-like system. The remaining $\sim 40\%$ are wider detached BH-star binaries. 

\section{Conclusion and Discussion}\label{sec:diss}
The detection of stellar-mass black holes through electromagnetic signatures (e.g., X-ray binaries), gravitational waves from merging binaries, and astrometric measurements of wide-orbit companions (Gaia BHs) has revealed a diverse population of systems. While each observational method probes a range of different physical processes and possibly stages in the life of a black hole, connecting these distinct populations has remained an open challenge. 

Here,  we propose a new formation channel that naturally bridges these populations. By focusing on massive stellar triples, which constitute the majority of massive-star systems \citep[e.g.,][]{Moe+17}, we show that the merger of an inner black hole binary, followed by the resulting recoil kick, can lead to four distinct outcomes:
\\ \noindent
{\bf (1)} a prompt transient electromagnetic counterpart to the gravitational wave merger due to the tidal disruption of the star by the black hole and subsequent highly super-Eddington accretion.  The transient will occur about 10 days after the BH merger (Fig. \ref{fig:TimeDelay}) and will likely be a bright optical-UV flare lasting a few days to a week, in some ways analogous to luminous fast blue optical transients.  In a subset of cases, the transient will be a partial tidal disruption leading to a repeating electromagnetic counterpart with a period of order two months.
\\ \noindent
{\bf (2)} a black hole–stellar system that slowly circularizes due to tidal interaction, eventually undergoing mass transfer and producing a low-mass X-ray binary (likely Gyr after the gravitational wave merger).
\\ \noindent
{\bf (3)} a wide black hole–stellar companion system, akin to Gaia BHs; or
\\ \noindent
{\bf (4)} an unbound system (i.e., a single, isolated BH).
\\ \noindent These outcomes are illustrated schematically in Figure~\ref{fig:cartoon} and through examples in Figure~\ref{fig:OrbitEffectExample}.   An important feature of our model to stress is that the tertiary companion is not dynamically important in the evolution of the inner binary.   Instead, we have explored the outcome of a passive tertiary in an otherwise binary-driven black hole merger channel.  
%(1) a transient source, such as a super-luminous even or an X-ray binary. On such an outcome, the BH is either crossing the star's tidal or Roche radius, resulting in a gravitational wave followed by an electromagnetic signature, with a short time delay (about two months on average) (2) a wide black hole–stellar companion system, such as the Gaia-BH systems  or (3) an unbound configuration (i.e., a lonely BH). See Figure \ref{fig:cartoon} for an illustration, and Figure \ref{fig:OrbitEffectExample} for two example systems.

To explore this scenario, we performed a proof-of-concept population study of BH binaries with a $1$~M$_\odot$ tertiary companion, motivated by the isolated binary BH merger channel \citep[e.g.,][]{Belczynski+02,Belczynski+07}. In this initial exploration, we fixed the tertiary mass and found that the final configuration depends most strongly on the star’s initial orbital separation, rather than other orbital parameters such as mutual inclination. Specifically, while counterintuitive, the mutual inclination does not play a key role because the new BH-star binary relative velocity is a combination of the kick and the pre-kick velocity.  As expected, wider-orbit stars are more likely to become unbound. Among systems that ended up in a close post-kick encounter (crossing the tidal radius), the pre-kick separation averaged $\sim 13$~au, with a wide tail extending up to $\sim 10^3$~au.

{In this proof-of-concept analysis, we adopted negligible BH natal kicks.  However, various theoretical supernova models predict significant mass ejection for BH progenitors across a wide range of parameters, yielding a wide range of kick velocities \citep[e.g.,][]{Fryer+12,Repetto+12,Muller+16,Mandel+20,Maltsev+25}. These kicks, along with the mass loss that occurs suddenly during BH formation \citep[e.g.,][]{Blaauw61,Hills83},can impact the configuration of triple systems and may even lead to the unbinding of the system. For instance, it has been suggested that such kicks could reduce the BH merger rate by a factor of a few \citep[e.g.,][]{Silsbee+17,Antonini+17}. On the other hand, some recent observational systems, including VFTS 243 and V404 Cygni, indicate negligible BH natal kicks \citep[e.g.,][]{Shenar+22,Vigna-Gomez+24,Vigna-Gomez25,Willcox+25,Burdge+24,Shariat+25xray}. Nonetheless, the uncertainty surrounding these estimates is considerable, and we caution that the rates discussed in this letter may have substantial uncertainties. The full evolution of triple stars during their lifetime, including BH natal kicks, is reserved for future studies.}

A key factor governing the outcome is the angle $\theta$ between the star’s pre-kick velocity vector and the kick direction. When the kick velocity is comparable to the stellar orbital speed, %retrograde configurations, 
where the vectors are aligned, are more likely to result in a small impact parameter, and thus, in close approaches. This follows from Eq.~(\ref{eq:Rc}), which shows that close encounters require small impact parameters relative to $G(m_\star + M_{\rm new})/v_{\rm new}^2$. Since eccentric companions spend most of their orbit near apocenter, close approaches typically occur when the kick velocity nearly cancels the orbital motion (see Section \ref{sec:outcomes} and the bottom panel of Figure \ref{fig:population}). For the parameter choices adopted here, we estimate that $\sim$0.02\% of LVK sources may be followed by an electromagnetic counterpart, with a short time delay (averaging around 10 days; see Figure \ref{fig:TimeDelay}).

Interestingly, about 13\% of our sample resulted in a bound star–BH configuration after the kick. Notably,  $\sim 8\%$ of the sample formed systems with post-kick orbital periods less than 10 years, comparable to the Gaia BH candidates. Our channel can resolve the difficulty of producing systems like Gaia BH1, where the stellar companion currently resides at a separation that would have been engulfed during the BH progenitor’s red giant phase. This tension has led to several proposed explanations, including fine-tuned mass transfer or envelope ejection, triple-star evolution, and natal BH kicks that reshaped the orbit \citep[e.g.,][]{El-Badry+23GaiaH1,Generozov+24,Li+24,Fishbach+25}. While the natal kick explanation shares features with the mechanism we propose, recent detections of X-ray binaries with tertiary companions suggest that at least some BHs may receive small or negligible natal kicks \citep[e.g.,][]{Burdge+24,Shariat+25xray}.  Our results show that Gaia BH–like systems can arise naturally following a black hole merger and recoil kick, without requiring extreme assumptions about natal kicks. Based on our toy model, we estimate that this channel could contribute between 1–10\% of all Gaia BHs in the Galaxy.

Crucially, the framework proposed here not only accounts for the gravitational-wave detections by LVK but also offers a viable pathway for forming Gaia BHs and, in some cases, X-ray binaries, thereby linking all three observational channels for the first time within a single evolutionary scenario.  It also predicts a robust channel for producing electromagnetic counterparts to binary black hole mergers.

\acknowledgments
We thank the anonymous referee for useful and detailed report.  
S.N. acknowledges the partial support of NSF-BSF grant AST-2206428 and NASA XRP grant 80NSSC23K0262, as well as Howard and Astrid Preston for their generous support.
Z.H. acknowledges support from NASA grants 80NSSC22K0822 and 80NSSC24K0440.  EQ thanks the Gordon and Betty Moore Foundation for support through grant GBMF5076.

\appendix

\section{GW Recoil Kicks}\label{sec:kicks}
Here we provide the fitting formula from \citet{Lousto+10,Lousto+12} for the post-merger recoil kick. We note that in our case, the velocity is in the plane of the binary. The general kick velocity vector is:
%which provides the recoil kick velocity within the plane of the binary BHs. Specifically, the recoil kick is:
\begin{equation}
    {\bf v}_{\rm kick} = v_m \hat{e} + v_\perp (\cos\xi\hat{e}+\sin\xi\hat{h}) +v_{\parallel}\hat{h} \ , 
\end{equation}
where $\{\hat{e}_\bullet,\hat{q}_\bullet,\hat{h}_\bullet\}$ is the Runge-Lenz coordinate of the BH binary, and the $\perp$ and $\parallel$ are the components that are perpendicular and parallel, respectively, to the binary's angular momentum, where $\hat{h}={\bf L}_{\rm BHs}/L_{\rm BHs}$
%(similarly,  $\{\hat{e}_\star,\hat{q}_\star,\hat{h}_\star\}$ is the  Runga Letz coordinate of the star about the BH's center of mass). 
Additionally, 
\begin{eqnarray}
v_m &= & A \eta^2 \sqrt{1 - 4 \eta} (1 + B \eta) \ , \\
v_{\perp} &=& \frac{H \eta^2}{1 + q} (\chi_{2, \parallel} - q \chi_{1, \parallel}) \ , \\
v_{\parallel} &=& \frac{16 \eta^2}{1 + q} [V_{1,1} + V_A \tilde{S}_{\parallel} + V_B \tilde{S}_{\parallel}^2 + V_C \tilde{S}_{\parallel}^3] \nonumber \\
&\times& |S_{2, \perp} - q S_{1, \perp}| \cos (\phi_\Delta - \phi_1) \ ,
\end{eqnarray}
where $\eta = q/(1+q)^2$ is asymmetric mass ratio, $\vec{\chi}_{1}$ and $\vec{\chi}_{2}$ are the dimensionless spin vectors, and the vector $\tilde{\bf S}$ is defined as
% Note that S_i is my chi_i.
\begin{equation}
\tilde{\bf S} = 2 \ \frac{{\bf S}_{2} + q^2 {\bf S}_{1}}{(1+q)^2} \ ,
\end{equation}
$\phi_1$ is the phase angle of the binary, and $\phi_\Delta$ is the angle between the in-plane component of
\begin{equation}
\vec{\Delta} = M^2 \ \frac{\vec{\chi}_{2} - q \vec{\chi}_{1}}{1+q} \ ,
\end{equation}
and the infall direction at the merger. Following \citet{Lousto+12} we assume that the angle $(\phi_\Delta - \phi_1)$ is chosen from a uniform distribution $[0, 2 \pi)$, and the constants are: $A = 1.2 \times 10^4 \ \rm{km} \ \rm{s}^{-1}$, $H = 6.9 \times 10^3 \ \rm{km} \ \rm{s}^{-1}$, $B = -0.93$, $\xi = 145^\circ$, $V_{1,1} = 3678 \ \rm{km} \ \rm{s}^{-1}$, $V_A = 2481 \ \rm{km} \ \rm{s}^{-1}$, $V_B = 1793 \ \rm{km} \ \rm{s}^{-1}$, and $V_C = 1507 \ \rm{km} \ \rm{s}^{-1}$ \citep{Gonzalez+07, Lousto+08, Lousto+12}. This model aligns well with full numerical relativity results, even in the intermediate mass ratio regime of \( q \sim 0.1 \) \citep{Gonzalez+09}. We also calculate the post-merger mass $M_{\rm new}$ using Equation~4 in \citet[][omitted here to avoid clutter]{Lousto+10}. 

%It will be supereddigton - see: 
%https://iopscience.iop.org/article/10.3847/1538-4357/add158/pdf
\bibliographystyle{hapj}
%\bibliography{papers2}
\bibliography{Binary}
\end{document}